# Geopolitical Implications of a Successful SETI Program

Jason T. Wright[1,2], Chelsea Haramia[3], Gabriel Swiney[4,5]

## Abstract:

We discuss the recent "realpolitik" analysis of Wisian & Traphagan (2020) of the potential geopolitical fallout of the success of the Search for Extraterrestrial Intelligence (SETI). They conclude that "passive" SETI involves an underexplored yet significant risk. This is the risk that, in the event of a successful, passive detection of extraterrestrial technology, state-level actors could seek to gain an information monopoly on communications with an extraterrestrial intelligence. These attempts could lead to international conflict and potentially disastrous consequences. In response to this possibility, they argue that scientists and facilities engaged in SETI should preemptively engage in significant security protocols to forestall this risk.

We find several flaws in their analysis. While we do not dispute that a realpolitik response is possible, we uncover concerns with Wisian & Traphagan's presentation of the realpolitik paradigm, and we argue that sufficient reason is not given to justify treating this potential scenario as action-guiding over other candidate geopolitical responses. Furthermore, even if one assumes that a realpolitik response is the most relevant geopolitical response, we show that it is highly unlikely that a nation could successfully monopolize communication with ETI. Instead, the real threat that the authors identify is based on the *perception* by state actors that an information monopoly is likely. However, as we show, this perception is based on an overly narrow contact scenario.

Overall, we critique Wisian & Traphagan's argument and resulting recommendations on technical, political, and ethical grounds. Ultimately, we find that not only are Wisian and Traphagan's recommendations unlikely to work, they may also precipitate the very ills that they foresee. As an alternative to the Wisian & Traphagan recommendations, we recommend transparency and data sharing (which are consistent with currently accepted best practices), further development of post-detection protocols, and better education of policymakers in this space.

[1] Department of Astronomy & Astrophysics; Penn State Extraterrestrial Intelligence Center; Center for Exoplanets and Habitable Worlds, 525 Davey Laboratory, Penn State University, University Park, PA, 16802, USA
[2] Corresponding author, astrowright@gmail.com
[3] Department of Philosophy, Spring Hill College, 4000 Dauphin St., Mobile, AL, 36608, USA
[4] National Aeronautics and Space Administration (NASA), Washington, DC, USA
[5] Harvard Law School, Cambridge, MA 02138

# 1. Introduction:

## 1.1 SETI and METI

The modern Search for Extraterrestrial Intelligence (SETI) was catalyzed when Cocconi & Morrison (1959) [1] performed a startling calculation. They showed that humanity's rapidly improving radio broadcasting capabilities were capable of producing signals just powerful enough that humanity's rapidly improving radio receiving capabilities could detect them across interstellar distances. This realization, and the work of Frank Drake to detect such signals from alien species at Green Bank Observatory under Project Ozma [2], led to a flurry of theoretical and practical activity to search for signs of intelligent life in the universe. The field of SETI today is broad, and encompasses any such search, including the search for any sign of non-communicative technological life. Such signs are called *technosignatures* [3,4].

In 1974, the Arecibo Message was composed and transmitted from the Arecibo radio telescope. This powerful transmission was directed at a cluster of stars thousands of light years away, and contained a short message describing the Earth, humanity, and the system used to generate the message. It was probably too brief to have any realistic chance of being detected; rather, it was "intended as a concrete demonstration that terrestrial radio astronomy has now reached a level of advance entirely adequate for interstellar radio communication over immense distances" [5]. Nonetheless, the message was the first modern attempt to send communications to an alien species, and the beginning of METI (for "Messaging Extraterrestrial Intelligence" [6], also called "active SETI"). Other examples of such proactive attempts at contact include the Pioneer Plaques [7], the Voyager Golden Records [8], and many radio messages similar to the Arecibo Message, mostly directed at nearby stars [9].

The details of and motivations for METI and "passive" SETI often overlap. For instance, the Soviet program was called CETI for "Communication with Extraterrestrial Intelligence" [10] and was just as concerned with how to establish 2-way communication and establish mutual intelligibility as it was with the actual detection of alien life. Zaitsev (2006b) [11] highlighted the symmetry between METI and SETI with the "SETI Paradox", which pointed out that in order for there to be signals for humanity to discover, there must be species willing to transmit them—if all species merely "listened" and did not "shout", then SETI would necessarily fail. Cortellesi (2020) [12] discussed a "continuum of astrobiological signaling" with METI being just the most intentional manifestation of the ways that Earth life might be detectable at interstellar distances via its biosignatures and technosignatures. For instance, Sheikh (2020) [13] performed a passive radio SETI search of stars in the Ecliptic Plane, because lifeforms on planets orbiting those stars would be able to easily detect Earth (and, probably, evidence for life it its atmosphere) as Earth passes in front of the Sun once per year from their perspective.

The implications of success for SETI and METI are assumed by many to be profound, and understandably so. Confirming scientific evidence of other intelligent life would be a truly novel and important discovery. There is significant study and debate about whether they might be detrimental to humanity.

## 1.2 Concerns About SETI

The social implications of success of passive SETI were studied in a series of workshops sponsored by NASA in 1991–1992. The resulting "Billingham report" [14] concluded that "Detection of an ETI signal would spark intense widespread interest that would in turn prompt new technological advances and highlight a need for organized assessment of the discovery." Later efforts to study these implications have been more interdisciplinary in nature and include meetings of the Breakthrough Listen Making Contact workshop[6], the Society for Social and Conceptual Issues in Astrobiology, as well as other projects and venues.

Perhaps the most common popular concern is that contact with an alien species would necessarily end badly for humanity by analogy with certain human contact scenarios, for instance because of culture shock in communicating with an "advanced" civilization, and because such contact might fuel "millennial enthusiasm or catastrophic anxiety" [14].

While the Billingham report acknowledged these possibilities, it also warned against leaning too heavily into historical analogies, noting that many human contact scenarios have been mutually beneficial, and the fact that "there are no historic events that are exactly analogous to an ETI detection…mandates caution in seeking insights into likely human reactions to such a detention from partially analogous historic encounters and events." The report concludes that there is substantial work to be done to determine what human reactions to a detection might be like, and that reactions would likely be highly variable across the world.

## 1.3 Concerns about METI

METI provokes much stronger concerns in some quarters because it seeks not just to discover alien life but to make such discovery mutual, inviting the possibility that the alien species could send us messages specifically, or even visit the Earth. Concerns that such contact could be extremely detrimental or even catastrophic to humanity are doubtless inspired, or at least amplified, by popular imaginings of alien invasions from science fiction (see, e.g., [15]), but are also driven by the historical analogies the Billingham report warned against. (To give a modern example, the physicist Stephen Hawking worried that "If aliens visit us, the outcome would be much as when Columbus landed in America, which didn't turn out well for the Native Americans. We only have to look at ourselves to see how intelligent life might develop into something we wouldn't want to meet.") Another popular concern involves questions about who has the right to speak for all of humanity [16,17,18].

---

[6] In fact, the 2018 meeting of the Breakthrough Listen Making Contact workshop led to an issue of the *American Indian Culture and Research Journal* (vol. 45, issue 1), in which, among valuable recommendations, is the call for even greater interdisciplinary work from the humanities and social sciences on debates in SETI and space exploration.

These concerns have given rise to a significant debate in the literature about whether METI should be proscribed at the governmental level. METI advocates and apologists argue, for instance, that our bio- and technosignatures are already so apparent that the practice invites no additional harm [19]; that the chance of harm is so unknowable that it is essentially impossible to decide how to treat it [20]; and that METI ancillary benefits outweigh the tiny risk of harm [9,21]. METI opponents range from those merely suggesting that message composition be careful not to anger, bait, or give too much information to potential species ([22,23], but see Santana 2021 [24] who recommends belligerent messaging); to those suggesting a moratorium until global consensus is reached [25,24]; to those who declare any such transmission "unwise, unscientific, potentially catastrophic, and unethical" [26].

## 1.4 Current State of Play

Today, there are no significant restrictions on SETI or METI beyond a general lack of governmental funding for either. Passive SETI projects are generally privately funded (the Breakthrough Listen project has a budget of millions of US dollars per year), although some smaller amounts of funding at NASA have begun to flow and the government-funded FAST telescope in China has SETI as one of its science goals. METI activities are rare and privately funded.

The SETI community continues to develop "post-detection protocols" for managing a potential detection. In 1989 the International Academy of Astronautics (IAA) Permanent Committee on SETI approved a Declaration of Principles Concerning the Activities Following the Detection of Extraterrestrial Intelligence (the First SETI Protocol, available in [14]). The Protocol recommends that researchers with a potential signal verify it and share its details with other researchers for confirmation prior to making any public announcements. If confirmed, researchers are then to announce the detection to various international bodies such as the International Astronomical Union, and the announcement should "be disseminated promptly, openly, and widely through scientific channels and public media." It also states that "no response…should be sent until appropriate international consultations have taken place." This Protocol has no binding force, but is widely known in the SETI community.

The Billingham report further recommended a range of studies and preparations that should take place to prepare NASA for a potential discovery (although these recommendations were to NASA, they are written generally and could apply to any entity that might handle such a detection.)

Since then there has been significant work on post-detection protocols, much of it under the auspices of the IAA Permanent Committee, which updated the First SETI Protocol in 2010[7], and continues to consider such matters as the public communication of detections (for instance via the Rio Scale, [27]) and the propriety of METI.

---

[7] Available at http://resources.iaaseti.org/protocols_rev2010.pdf. Retrieved March 4, 2022.

The topic continues to inspire healthy and active academic debate as well, as the instant paper illustrates. (See also [28,29,30] and references therein, the last of which highlights the potential for social and political disruption post-contact).

## 1.5 Wisian & Traphagan

Wisian & Traphagan (2020) [31, hereafter W&T] recently undertook "a realpolitik consideration" of the dangers of passive SETI, and concluded that SETI could be as risky as METI in certain respects. Their focus is not on the risks that the ETI themselves might pose, but on the utterly terrestrial issue of how humans might predictably react at the state level to a confirmed detection of extraterrestrial intelligence.

W&T propose specific measures in light of the geopolitical risks they highlight. Their recommendations for scientists engaged in SETI and METI include "supporting relationships with local and regional law enforcement organizations," "physical/perimeter security" of their research facilities, and "personnel security" for them and their families. They further recommend that facilities used by these researchers, such as radio telescopes, adopt other security measures akin to those at nuclear power plants or biowarfare laboratories.

# 2. Statement of the Wisian & Traphagan Premises

W&T reason in the following way. In a post-detection scenario, humans themselves present a threat to humanity. This threat comes from the possibility that state leaders will react according to realpolitik motivations. Quoting Bew (2016) [32], W&T define 'realpolitik' as "the view of interstate relations where '*the notion that the state could be regulated or controlled by law [is] flawed' and that 'power obey[s] only greater power.*'" According to this articulation of the realpolitik worldview, national interests are necessarily focused on domination and on acquiring and maintaining power. Historical examples demonstrate that realpolitik motivations can quickly lead to conflict when there is the possibility for power or control over valuable assets or resources.

Working within this realpolitik framework, W&T claim that communication with or information from ETI could be viewed as highly valuable. They are worried that nations could therefore attempt to exercise their power over other nations in order to monopolize control of transmissions and communications, leading to the possibility of an information arms-race. This race could lead to war or other disasters, and it could put scientists and their facilities under direct threat as well.

Crucially, the opportunity for seizing control of communications need not be realistic. The mere perception of the opportunity for a power-grab could be enough to jump start an information arms-race, and therefore nations need not be justified in their attempts to control communication for those attempts to raise the concerns that W&T outline. In tandem, states who lack sufficient interest in securing power over ETI communications could nonetheless perceive that other states will have that interest, and therefore perceive (rightly or not) that they are under a national security risk from those states.

Amidst these possibilities and perceptions is the crux of W&T's worries. They argue that there is good reason to conclude that successful detection could lead many nations to react according to their own or others' realpolitik motivations, whether those motivations are real or merely perceived. Thus, there is good reason to implement extreme security measures in anticipation of the above risks.

We refer below to this concern that state actors will act in this way, justifying the W&T recommendations, as "the W&T realpolitik scenario". This is what W&T argue has a sufficiently high probability of happening to guard against.

We refer to the specific contact scenario that would trigger the realpolitik response as "the W&T contact scenario." It is not essential to the W&T realpolitik scenario that this contact scenario be likely, even in the event of contact: in principle, the mere perception it has occurred is sufficient to trigger their realpolitik scenario.

## 2.1 Unpacking the technical argument

The W&T contact scenario involves a message which we can respond to and which contains significant intelligible information, including new scientific knowledge. Because a technologically capable alien species will almost undoubtedly be much older than modern human civilizations, W&T argue that "even the most seemingly trivial of resulting scientific knowledge, if wielded solely by one nation here on Earth, might enable it to dominate the world." They imagine, presumably, something along the lines of new engineering or physical insights that could enable the production of a weapon or technology so much more powerful than anything in current arsenals that it could allow any state to dominate all of the others.

W&T also claim that unless the signal comes from a transmitter within the Solar System (that is, if it is at interstellar distance) the equipment needed to send and receive signals is of the scale of a major radio observatory, of which there are "around a dozen" in the world. That is, they assume that the primary difficulty in engaging in contact with the species will be controlling one of a small number of large, expensive, and sophisticated facilities (such as the 100-m Robert C. Byrd Telescope in Green Bank, West Virginia, USA, or the 500-m FAST telescope in Guizhou, China)

A second assumption they make about distant alien species is that, without "local observers/probes," they will have no way to "independently verify what Earth tells it." Their meaning here is not entirely clear, but they apparently envision that a nation will see a strategic advantage in being able to tightly control what the species learns about us, and that despite the species' great technological advantage, they will not be able to confirm what they are told by seeing details of humanity's activities on Earth's surface or detecting our weaker transmissions.

## 2.2 Unpacking the political argument

W&T note that, since the 1989 Protocol is aspirational, voluntary, and only applies to scientists, there are no enforceable regulations or protocols to ensure that nations collaborate when presented with the opportunity to communicate with ETI. So, they argue, once the success of METI or SETI creates the option of communicating with ETI, it is likely that realpolitik motivations will take over: nations will fight each other for control over the communication and information, leading to harmful and potentially catastrophic effects on humans and on human civilization.

Again, W&T define realpolitik as "the view of interstate relations where '*the notion that the state could be regulated or controlled by law [is] flawed*' and that '*power obey[s] only greater power*'" [32]. They acknowledge that although other ("liberal") models "do play a significant role in international relations, the…realpolitik model continues to be an important component of how governments behave and react when faced with concerns over potential threats from outside forces" and that "calculations of national interest cannot be ruled out as potentially dominant in any situation that involves the possibility of competitiveness among governments, including an event as significant to humanity as a confirmed SETI discovery."

# 3. Narrowness of the W&T Scenarios

## 3.1 The contact scenario

The W&T contact scenario is very narrow, and is arguably an unlikely outcome of SETI work. This is because the postulated contact must have a very particular character: 1) the signal must be from one of the nearest stars, 2) communicative, 3) intelligible, and 4) information rich; 5) it must be strong enough to provide dense information content, but 6) weak enough that only the largest telescopes or telescope arrays can detect it; 7) a small number of exchanges must be sufficient to derive information about "new physics"; and 8) this new physics must be powerful enough to be translated into a dominating technology, but 9) it is not so "advanced" that we have no hope of quickly understanding and implementing it.

### 3.1.1 Likelihood of Communication

First, such contact is only possible under certain forms of SETI.

SETI searches for any technosignature, not just communicative signals intended for humanity. Out of necessity, early radio work focused on "beacons" ("loud," obvious signals intended to announce a species' presence, Wright et al. 2018), and much early speculation went into what information such a beacon might contain. Searches today, such as the Breakthrough Listen radio search (e.g. [33]) are sensitive to a wide range of signal strengths and waveforms, and are more agnostic regarding the motivations for and targets of the transmission. Other forms of communication SETI include searches for laser pulses (whether for communicative purposes or not).

But SETI is broader even than this and includes (for instance) searches for atmospheric technosignatures, or searches for the waste heat of industry. Discovery of life via these means would have a very different character than in the W&T contact scenario: these technosignatures are ambiguous, and as such would not, without additional lines of evidence, provide a definitive demonstration that technology had been discovered. The discovery of intelligent alien life—and any communicative signals they might be sending us—may thus be a long and continuous process, not a discrete moment that would trigger the W&T realpolitik scenario.

Nonetheless, W&T are correct that some forms of SETI seek, and may find, communicative signals. If the signal discovered were communicative, W&T maintain that governments could assign a high likelihood that the signal will be intelligible and contain non-trivial information. This could be true: especially if the signal is a beacon, it may be "anti-cryptographic" [34] and intended to be decoded by any intelligent species.

We note that the problem of how to generate such a message is a difficult one—the Arecibo Message was intended to have this property, but the consensus in hindsight is that it is too obscure to be understood even by other humans. Indeed, the problem of interspecies communication is a thorny one, and we have not even been able to translate messages from species with which we share significant evolutionary descent, such as dolphins, despite being able to study them closely and extensively. We simply do not know what a truly alien consciousness would find obvious or clear.

But to continue, let us grant that, despite these difficulties, the signal detected is communicative and intelligible, noting for now only that such a thing may be objectively quite unlikely.

### 3.1.2 The value of contact

W&T next claim that even the most trivial (to the aliens) facts about the universe contained in the message could be useful to us as scientific advances that can be translated into revolutionary technologies.

That this could happen is not obvious at all. First of all, science is cumulative and nonlinear: in order for a new insight to be useful, we must first have the appropriate scientific context to understand it. To use a (fraught) historical analogy, if medieval scholars had somehow been given a translation of a textbook on nuclear weapons design, it would be essentially useless to them. Even if they completely understood it (which they could not, lacking any background in nuclear physics, modern engineering, rocketry, etc.) they would not have the technical capabilities to do anything with it. In other words, unless we are already on the verge of a major scientific breakthrough, simply knowing that the breakthrough is possible is not going to precipitate it.

Secondly, it's not clear what sort of technological advantage W&T have in mind. Strategic nuclear weapons already present destructive force far beyond what is practical for war. Even more advanced weapons would be unlikely to destabilize the international system; a country possessing some alien weapons system would still be subject to nuclear deterrence. Other potential technologies — something approaching propulsionless drives, perfect cloaking, or teleportation perhaps — would seem to be very sophisticated technologies far beyond anything we could produce in the near future. It seems unlikely that there could be any simple morsel of physics not known to us that would provide the key final steps to allow a state to quickly develop such technology.

One could still worry that simply knowing that a breakthrough is possible would hasten our reaching that milestone, even if it didn't immediately precipitate it.[8] Suppose an alien message convinced us that technology that we believed to be impossible (e.g., faster-than-light travel, anti-gravity, etc.) was in fact possible. Notably, only one or more sufficiently powerful state leaders would need to be convinced of this for realpolitik actions to potentially follow—and they need not be convinced according to good evidence. To respond, we should separate the challenge that arises from the prospect of an *actual* breakthrough from the challenge that arises from the mere *perception* of a breakthrough.

Assuming there is a real breakthrough to be had, hastening the discovery of a scientific or technological breakthrough sounds more dramatic than it presumably would be. At least some developed countries are already engaged in the project of attempting to be the first to make certain breakthroughs. To speed up a breakthrough is not necessarily to make it happen quickly. Hastening a breakthrough in this scenario could simply mean that a breakthrough occurs in the next few centuries instead of the next millennium, but the likelihood of extraterrestrial information triggering a *sudden* breakthrough is still far lower than it simply hastening a breakthrough. Continuing such attempts in the event of successful detection might not change much, then, insofar as the most effective competitive edge doesn't come until the breakthrough itself arrives. And the sudden ones are the ones that are most likely to trigger the W&T realpolitik scenario.

But even if such a technology breakthrough could be quickly realized, implemented, and exploited, that would imply that all other nations would be similarly close to such breakthroughs, and that the deployment of the technology would trigger a similarly quick adoption of the technology worldwide. So even in this optimistic scenario, the monopoly would be short lived.

So, an actual competitive edge is still unlikely to arise immediately post-discovery, especially a long-lived one. However, this implausibility will not necessarily prevent state actors from misperceiving the possibility of an immediate competitive edge—a misperception that could lead to attempts to secure an information monopoly. The attempts themselves could trigger the W&T realpolitik scenario, and such attempts are not required to be grounded in good science or good reasoning. In response, we contend that the best way to combat potentially catastrophic misperceptions is to make the good science and good reasoning as clear as possible to leaders and others who may be susceptible to misperception. We will return to this point later.

---

[8] Thanks to Ian Stoner for pointing out this concern.

It is nonetheless conceivable that there might be certain edge cases, where our current level of understanding is sufficient to utilize some ET-imparted insights, such as with fusion technology or advanced artificial intelligence, that could confer profound and lasting advantages on the recipients. Similarly, some countries may wish to play the long-game and attempt to monopolize the possibility for a breakthrough that they reasonably believe will not occur for centuries or longer. However, these areas of research are by their nature *ones that humanity is already engaged in separately from SETI/METI efforts*. Given that fact, it is possible—and perhaps likely—that humanity would develop these technologies eventually anyway, regardless of ET intervention. More importantly, it is notable that the international community has responded to the danger of monopoly power in these areas not through realpolitik, but by focusing on international collaborative efforts and developing widely-adopted best practices for information-sharing and transparency.

For example, traditional rivals such as the United States and China have agreed, along with other countries, to cooperate together in the world's largest project to develop fusion power technology—possibly the most game-changing technology currently under development through the International Fusion Energy Organization (ITER).  The rules governing ITER are set out in a multilateral treaty, and provide that "any scientific results shall be published or otherwise made widely available…"  (ITER Agreement, Article 10)  The ITER Agreement goes on to provide that every Party has a right to all intellectual property generated by the organization on an equal and non-discriminatory basis.  (Annex, Article 5)  This sort of open sharing of data is precisely what we recommend below. We note that the effectiveness or wisdom of agreements like ITER is beside the point, which is just that history shows that such agreements are a common avenue for managing potentially geopolitically disruptive technology breakthroughs.

### 3.1.3 Possibility of an information monopoly

W&T next assume, almost tacitly, that two-way communication will likely be possible and useful. If the transmitter is in the Solar System, this is correct, but W&T are concerned with transmitters that are "light years distant" and which cannot verify the information we send via nearby probes.

W&T claim that there are only a small number of facilities around the world that could provide access to this signal, citing Gertz (2017) [35, misspelled in their references]. But that work does not back up their assertion: indeed Gertz states that "…at least among Western democracies, hiding the coordinates [of a signal] would be difficult, given that numerous telescopes would (and should) be enlisted in the follow up confirmatory observations."

"Numerous" here is not "around a dozen." Modern radio telescopes are large and expensive because they are general purpose instruments, akin to a complete toolbox. They can often point in any direction and have a suite of specialized instrumentation designed to operate over a huge range of frequencies and timescales.

But once a signal is discovered, the requirements to detect it shrink dramatically. Only a single narrowband receiver is required, and the bandwidth of its receiving backend need be no wider than the signal itself. The telescope need only point at the parts of the sky where the signal comes from, so it need only have a single drive motor. And the size of the dish need not be huge, unless the signal just happens to be of a strength that large telescopes can decode but small ones cannot, which is possible but *a priori* unlikely. To extend the toolbox analogy: once one understands that the problem at hand only involves turning a screw, one only needs to acquire a screwdriver.

Indeed, there are an enormous number of radio dishes designed to communicate with Earth satellites that could easily be repurposed for such an effort and could even be combined to achieve sensitivities similar to a single very large telescope, if signal strength is an issue. And there is no shortage of radio engineers and communications experts around the world that can solve the problem quickly and easily. The scale of such a project is probably on the order of tens of thousands to millions of US dollars, depending on the strength and kind of signal involved. The number of actors that could do this worldwide is huge. Also, such efforts would be indistinguishable from normal radio astronomy or satellite communications, so very hard to curtail without ending those industries.

The situation is similar for a communicative pulsed laser signal: then the difficulty is primarily in the commercial availability of very fast detectors (e.g. [36]). Here, the technology is not as mature, and if the pulses are *extremely* fast (nanosecond scale) or at wavelengths not commonly used in commercial hardware then it is possible that the necessary technology could be controlled in the short term, but in that case the signal is unlikely to have been detected in the first place. As with radio, there are an enormous number of optical telescopes which will have similar sensitivity to optical flashes as existing optical SETI experiments (which, again, are expensive only because they search a huge fraction of the sky for signals of unknown duration).

But even if there are only a few, major installations capable of communication, the utility of such communication and the nature of a monopoly on those installations will be strongly mitigated by light travel time.

The nearest star (and nearest target of SETI searches beyond the Solar System) is 4 light-years away, meaning that any response to any signal we send will not arrive for *at least* 8 years (the light-travel time to the star, then back again). Unless this communication is extremely efficient and the useful information can be extracted in a single exchange ("Press 1 to download the *Encyclopedia Galactica*"), the conversation will occur over many decades, during which there will be ample opportunity for new facilities to come online throughout the world to join the conversation. Most targets of SETI searches are at least 200 light years away, meaning that the most likely outcome of W&T's scenario is that any communications from us will go unanswered for many human lifetimes.

This strongly limits the possibility of useful information exchange on timescales short enough that rival states cannot break the monopoly with new infrastructure of their own.

### 3.1.4 The Narrowness of the W&T Contact Scenario

Ultimately, then, our current science indicates that a contact situation could proceed in a variety of ways not acknowledged by W&T. Thus, we have highlighted the ways in which many contact scenarios could easily diverge from the scenario that W&T envision. Notably, we are not arguing that any *particular* contact scenario is likely. Rather, the point is that the range of contact scenarios that follow the outline W&T suggest is quite narrow, and only in W&T's narrow range of scenarios are their recommendations relevant.

We acknowledge that the W&T contact scenario is not appearing in a vacuum—it is a quite common popular perception of the outcome of SETI. Indeed, it shares many elements of the plot of the novel and film *Contact* by SETI proponent Carl Sagan [37,38], and the novella "Story of Your Life" [39] and its popular film adaptation *Arrival* [40]. Many other SETI practitioners have outlined why such a contact scenario may be likely, or expressed optimism that a communicative signal is likely to be found in decades, or expressed hope the problem of intelligibility has been solved by the transmitters of the signal.

We do note, however, that one should not necessarily take all of these statements at face value. When speaking with the public—and especially with potential funders—SETI practitioners naturally emphasize the most interesting and beneficial *possible* outcomes of their work. We should not confuse such advertisements and optimistic assessments for sober analyses of likely outcomes. *Contact* was a work of popular fiction, after all, designed to dramatize SETI to cinematic proportions.

Regardless, the narrowness of the W&T contact scenario and the mere perception of their probability by state actors both have important implications for the relevance of their recommendations. Because those implications follow from their political argument as well, let us first outline some concerns with the narrow applicability of their political premises.

## 3.2 The political scenario

The W&T realpolitik scenario is based on certain assumptions about the political nature of humanity. These assumptions deserve further scrutiny as well.

As W&T admit, "*Realpolitik is a controversial term within the international relations community…The term is subject to widely disparate definitions and has even been used derogatorily at times.*" Despite this caveat, W&T adopt—and premise their analysis on—a specific interpretation of realpolitik. As we have described above, this interpretation reduces to the idea that "power [obeys] only greater power."  This understanding of international relations is flawed, leading to inaccurate implications for SETI.

First, international relations scholars within the realpolitik school of thought do not themselves agree on what motivates the behavior of states. While it is certainly true that many "realist" commentators consider the pursuit of power to be a primary motivator for states, most also accept that other considerations also play a role. For example, Hans Morgenthau, perhaps the most influential of all realist commentators of international relations, accepts that moral considerations must also be taken into account through a filter of "prudence" that considers political consequences *together* with moral considerations. [41] A range of other scholars accept that pursuit of power drives international relations, but differ in the reasons that states seek power: contrast the "defensive realism" views of Kenneth Waltz with the "offensive realism" of John Mearsheimer. [42,43] These differing motivations could have very different implications for how states would react to detection of communications from an ETI.

Although a detailed rebuttal of realpolitik and the realist school of international relations is beyond the scope of this article, the fact remains that realists themselves do not agree on what motivates state behavior. More to the point, W&T simply ignore this ongoing debate, focusing instead on what is essentially a caricature of realism in which states pursue power apparently for its own sake, devoid of other considerations or goals.

Put simply, there is no reason to assume—as W&T do—that states will behave in the way they assume. It is of course possible. However, other candidate (non-realpolitik) responses seem equally likely. Other theories about the nature of international relations—and the corresponding anticipated geopolitical responses—would generate different prescriptions. Sufficient reason is not given to treat the possibility of a realpolitik response as action-guiding over other candidate responses that would justify different or even conflicting courses of action. As many authors have noted, states are guided by many considerations other than power in their decision-making, particularly in the fields of science and technology.  (See [44] for a discussion of the relevance of commercial factors, [45] for an examination of the unifying nature of technical endeavors, and [46] for a complete overview.)

For example, a competing account involves the claim that the pursuit of influence or prestige (rather than the pursuit of control) better describes the motivations of many actual nations. States can be motivated by power, of course, but influence, prestige, and other less quantifiable factors can also play a significant role. Even in the case of China—the state perhaps most likely and capable of behaving as W&T fear—the government places extraordinary emphasis on securing domestic and international prestige, and not only on securing technological successes. [47,48][9]  Likewise, recent experience during the Covid-19 pandemic demonstrates that states are willing to give away some hard-power advantages, in the form of vaccine technology and medical equipment, in the pursuit of influence.  (For a detailed examination of China's "Covid

---

[9] China Power Team, "What Is the Source of China's International Prestige and Influence?", China Power Team, China Power. July 27, 2016, Updated August 26, 2020. https://chinapower.csis.org/source-chinas-international-prestige-influence/. Accessed March 4, 2022.

Diplomacy" see [49,50]). Applied to the question of SETI and METI, the limitations of W&T's predictions are clear. They assume that the physical benefits of contact will be, or might be believed to be, so overwhelming, that a state *would actively suppress details proving that that they discovered alien life*, thus foregoing the incalculable prestige that would accrue to the state that made arguably the most significant scientific finding of the modern era. As with many of W&T's assumptions, this outcome is not impossible. However, they have not shown that this possibility is more likely than the possibility that nations will *avoid* pursuing an information monopoly insofar as those states are motivated not by domination but by the benefits of prestige or influential collaboration.

It is also relevant to consider the claim that international relations have followed an increasingly peaceful trajectory, and a growing number of nations form a group for whom conflict between any two of those nations seems absurd, despite significant power differentials. As philosopher Michael Huemer notes,

> Those who seek to explain international relations in terms of power relations and who emphasize deterrence as a necessary condition for security must have difficulty accounting for the continued enjoyment of peace and independence by…defenseless nations. [51]

Huemer notes that democracies tend to be part of the conflict-free group, whereas dictatorships tend to face conflict. Of course, the global increase in democratic values is not the only potential explanation for this increase in peace. Another viable explanation, suggested above, is that it is reliably in a nation-state's interests to aim for the benefits of (some combination of) collaboration, influence, and prestige, and an apparent commitment to peace may be an important part of securing these benefits.

Of course, one may reject claims about the reality of increasingly peaceful relations on the grounds that they are overly optimistic or inaccurate. Peace and cooperation between nation-states may be a fluke or a passing trend rather than an increasingly common outcome. Or, it may accurately describe a particular set of international relations, but not humanity's political trajectory on the whole. There is nonetheless a larger point that remains in the face of such challenges. Political attitudes unquestionably shift, develop, and remain unpredictable to a certain extent. And we must take seriously the facts that political attitudes are mutable, that global politics will change, and that realpolitik motivations are not always reducible to simple desires for power. We should not assume that most or all historical political relations–including realpolitik responses–are straightforward, static, or inevitable. Policy should reflect this propensity for change.

# 4. Evaluating the appeal to collective self-interest

In addition to technical and political concerns with W&T's argument, we may also explore certain of their ethical assumptions. W&T invoke the distinction between self-interest and altruism when constructing their argument, appealing to self-interest as the driving factor in realpolitik responses. They write "in other words, realpolitik can be thought of as placing collective (governmental) self-interest ahead of ideological or altruistic priorities in international affairs."

W&T assume that collectives can have interests, and they ask us to consider the collective interests of a government or nation. In the W&T realpolitik scenario, self-interest and altruism compete with one another, and nations have a self-interest in monopolizing communication in a contact situation rather than self-interestedly or altruistically sharing information.

One quick rebuttal to this involves the appeal to the ways in which collaboration—rather than monopolization—could reliably satisfy the self-interested motivations of individual nations. If the concern is that nations will act out of self-interest, then we should note that the W&T realpolitik scenario doesn't properly capture the scope of self-interested behavior. Their scenario is insufficient or inaccurate whenever national self-interests extend beyond or fail to motivate power grabs or domination and instead produce the motivation for collaboration and the like. States may self-interestedly wish to garner prestige or influence—or they may even have an altruistic interests in peace and cohesion—and they could recognize that, in many instances, these outcomes are best achieved through collaboration rather than domination. Put simply, we should not overlook the ways in which altruistic or collaborative motivations can satisfy the interests–including self-interests–of many nations, and we should note that these behaviors run counter to the W&T realpolitik scenario.

But there is another rebuttal to consider here. First, if we assume that interests can manifest at the level of a collective, then there seems to be no reason why collective interest boundaries must be no larger than those of nation-states. There could be broader collectives, the most relevant candidate in a contact scenario being the global-collective of humanity. And the collective of humanity may have an interest in avoiding individual nation's attempts at power grabs or information monopolies.

One may object at this point by claiming that this is not a legitimate collective because 'humanity' doesn't rigidly designate anything. While there are lots and lots of people, they have many and varied interests and little to unify them under a single referent.[10] However, the same charge can be brought against reference to a 'national' or 'governmental' self-interest. Nations are composed of lots of people with many and varied interests, and individual interests need not and often do not align with what is put forward by the government as national self-interest. Yet, this does not entail that legitimate collective interests are impossible. Nations have a shared history or shared territory that could justify the appeal to collective national interest.[11] Likewise, humans share a planet and a biological history that could justify an appeal to collective self-interest in the face of the diversity of interests of individual members of humanity. Claims of collective interests might not be straightforward or uncomplicated, but they might also be legitimate.

---

[10] Thanks to Kathryn Denning for raising this important concern.

[11] We do not mean to claim that a shared history or shared territory is necessary or required for justifying reference to a bloc of collective interests. We simply argue that shared histories or territories can be sufficient to justify reference to a bloc of collective interests, and other qualifications may also be sufficient.

However, the important worry here—assuming that collectives have legitimate interests—is that those who act on behalf of a collective could erroneously present those actions as collectively self-interested when they are not. In fact, in many instances, what has been presented as collective, national self-interest has actually been *un*representative of the national collective in question and instead merely represented the interests of those few who control the resources or military power of that nation. These instances are more than mere counterexamples; they are evidenced in realpolitik scenarios throughout history. While the actions of those in power can and often do affect everyone in the respective nation, it can also be a mistake to present such actions as representative of the nation's collective interests. What a national collective has interest in is not always represented by those who act on behalf of that nation. Notably, W&T refer to the collective in question as a "governmental" one, and in doing so seem to recognize that the power to act on allegedly national interests rests on the nation's leaders and not necessarily on the collective itself.

There are two lessons to take here. One is to be careful to note that exercises of power by governmental or national leaders can at the same time fail to represent the national collective and could even involve abuses of power. To present those actions as motivated by national self-interest is to erase or ignore true collective self-interest and to reinforce the powers that suppress those interests. This is not to deny that national leaders could still act from an interest in power or domination (nor to presume that collective interests could never include interests in power or domination); it is to caution against uncritically *presenting* the interests of a small cohort of disproportionately powerful individuals as coextensive with the collective's self-interests.

The above worry could be applied to virtually any appeal to collective self-interest, which leads to the second lesson. We should be careful to avoid framing collective-interest questions in ways that obscure the misrepresentation of the collective or legitimize abuses of power, and this includes appeals to the global collective of humanity. Thus, we should be careful to avoid conflating humanity's collective interests with, say, the interests of typically dominant or powerful members of humanity as we consider not only national collectives but also the prospect of the collective of humanity writ large.

With these worries in mind, let's assume, as W&T do, that a collective can act out of legitimate self-interest. A global collective would then reliably attempt to act in humanity's self-interest. While it is highly debatable what could or does qualify as legitimately serving humanity's self-interest, we should consider the possibility that communicating successfully with an extraterrestrial civilization could arguably be in humanity's self-interest for many reasons, given the potential benefits and utility involved. Conflict at the state-level would arguably inhibit humanity's ability to successfully communicate with an extraterrestrial civilization and could therefore go against humanity's self-interest. We are not assuming here that humans do or will have a collective interest in communicating successfully with ETI. The question of what humanity has a collective interest in deserves a far richer discussion than we have space for here.

The point, instead, is that, insofar as other collective interests—including and especially the interests of the collective of humanity, given that we are considering a contact scenario—could diverge from, conflict with, or override governments' purported national self-interest in monopolization of power, we must admit that other candidate collective interests are at least as relevant as the realpolitik interests that W&T focus on and should not be omitted from our analysis of how to mitigate risk of harm in a post-detection scenario, nor from the justifications of policy-recommendations.

# 5. Distinct concerns for METI

While W&T posit that SETI is as risky as METI, they overlook a distinct concern regarding METI that arises within their framework. On the one hand, METI actors are not simply searching; they are actively messaging, and they seek to elicit a reply. Successful METI then *entails* successful *communication*. Upon successful messaging, those METI actors in particular will, at least initially, have a monopoly on the communicative power that W&T worry will create conflict. SETI searches, on the other hand, set out to detect evidence of extraterrestrial technology. Successful SETI searches do not guarantee communication the way that successful METI searches do.

In the wake of a successful SETI detection, no one yet has the upper hand, and the question of who in particular will communicate, and what they say, remains open. In the wake of successful METI messaging, however, the messengers have already secured the role of communicator and determined the content of the (initial, at least) message. The point is that, even if we assume for the sake of argument that realpolitik concerns are as legitimate as W&T indicate, what realpolitik states will do when no one yet has the upper hand (a mere-detection SETI scenario) is presumably different from what they will do if one party already does (a successful-messaging METI scenario).

In fact, there is some reason to think that successful SETI activities are more likely to evoke a collaborative response than METI activities. States may wish to prioritize their potential interests in influence, prestige, or collaboration over other potential interests in domination or control precisely when there is no particular party to dominate or control. We do not presume that all states would undoubtedly be motivated in this way. Rather, we wish to highlight that METI by its very nature establishes a communication-monopoly before contact, whereas SETI allows for the possibility of the communicating party to remain open and thereby may avoid establishing a communication-monopoly before contact. In the former METI scenario, the power of communication is preemptively in the hands of the messaging party, which could lead to conflict regarding concerns about the legitimacy of that particular party's power.

Nonetheless, W&T are careful to note that their focus is on the risks of *communication per se* and not on mere detection. And, as they acknowledge, the FAQ section of the SETI Institute's website indicates that SETI practitioners are indeed open to the prospect of replying to a detected signal, and thus to initiating communication. It is in this way that SETI and METI may overlap (though a SETI scenario would presumably involve delay and deliberation before messaging, unlike in a METI scenario, thereby avoiding the distinct concerns with METI noted above).

However, W&T misrepresent the SETI Institute's justification for embracing the prospect of messaging. They state that "the SETI Institute in its frequently asked questions sections categorically states that there is no risk even from going beyond listening to actively messaging." Yet, the SETI Institute's FAQ section asserts something importantly different. To quote verbatim, it says

> …there's little point in worrying about alerting others to our presence by either deliberately transmitting (METI) or replying to a signal detected by SETI. That's because we have been unintentionally broadcasting the fact of our existence into space ever since the Second World War.

Notably, to say that there is "little point" in *worrying* about a risk is much different from saying that there is *no risk*. As we can see above, the SETI Institute does not appeal to a lack of risk to justify responding to a signal. Rather, they treat the risk—whatever it might be—as a foregone conclusion, given the likelihood that our presence is already detectable by virtue of our unintentional radio broadcasts. Claiming that something that might be harmful is very likely to occur regardless of what you do is a way of *acknowledging* the risk involved, not disputing it.

So, even where SETI and METI practices overlap, SETI practitioners are not as resistant to claims of risk as W&T indicate. They simply note that the risk is already present due to our own technosignatures. This risk has been debated, as noted above. Perhaps our unintentional broadcasting is too weak to be detected. Perhaps it is not. We can acknowledge that both are at least possible. But if we wish to take seriously the possibility that detection of humanity by ETI is a foregone conclusion, the following challenge can be applied to W&T's argument.

When presenting the risks of realpolitik responses, they consider only the agency of humans—specifically, our nations—ignoring the agency of ETI. They do not consider the possibility that the behaviors of Earth's nations and its international relations could affect our prospects of communication. Assuming that ETI could observe us to some degree, we might consider the possibility that ETI wishes to exercise control over whether and when they are detected by humans. It is possible that ETI may be waiting on evidence that we are worth communicating with, perhaps by our proving that we are capable of overcoming base, competitive attitudes and engaging in widespread collaboration. This suggests that realpolitik responses in general are risky in ways that W&T don't consider. The risks here would not arise from contact *per se*. Instead, there is a risk of preventing contact and thereby preventing the potential benefits of communication outlined by W&T. Of course, as they acknowledge, these benefits are speculative, as are the above proposed motivations and capabilities of ETI.

Nonetheless, these are worthwhile speculations for anyone who takes realpolitik concerns seriously, and they provide motivation to recognize the possibility that our conduct could conceivably affect the probability of contact, and, at the same time, they provide further justification for the claim that states could be motivated by the hope of beneficial communication with ETI to showcase better cohesion and open collaboration across humanity.

# 6. Shortcomings of the W&T recommendations

Because W&T view competition and attempts at monopoly to be likely in SETI/METI efforts, they suggest that "The SETI community should begin (if not already started) serious conversations on the ways in which lessons learned from [chemical and bio-weapon, nuclear, and abortion service facilities] could be employed at facilities such as radio telescopes." Although they do not provide detailed suggestions, the basic suggestion is clear: hardening of the information and physical security of SETI/METI facilities and practitioners.

Even assuming the dangers presented by W&T are likely and thus should be action-guiding, these security-focused actions are likely to fail or, worse, to create more problems than they solve. With regard to physical security, such efforts would only be possible even in principle for dedicated astronomy facilities. As we have described, any number of commercial and other facilities could be repurposed for SETI/METI efforts; nuclear-style hardening of these thousands of sites is not a realistic goal.

In nuclear, biological, and similarly secure fields, the actual material products of their work is the subject of a large security focus: radioactive materials and pathogens must not be allowed to fall into the wrong hands. With SETI and METI efforts, there is no physical product, and the only subject of possible competition is information itself. That fact means that if SETI and METI efforts were to become secure, information security would have to be the primary focus. We can look to history to remind ourselves that nuclear weapons, the best analogy to this situation, were a jealously guarded technology that nevertheless leaked via espionage within just a few years. Mathematical modeling by Grimes (2016) [52] demonstrates how hard it is to maintain secrecy over information for more than a few short years, which, as we have shown, is much shorter than interstellar communication timescales.

Implementing major information security measures as part of SETI and METI efforts would be extraordinarily difficult and possibly fatal to these programs. With the partial exceptions of Russian and Chinese efforts, most such efforts are collaborative.[12] For example, the Allen Telescope Array was largely funded by Paul Allen, was managed by the University of California Berkeley for a time, is now managed by the SETI Institute, and seeks additional funding from private donors and a range of governmental and scientific organizations. Optical SETI efforts are similarly collaborative, involving mixtures of academic, non-profit, private, and corporate funding and management. Major SETI facilities such as MeerKAT, the Byrd Telescope at Green Bank, and the Parkes Telescope are regularly used by astronomers from all over the world and are operated on behalf of university and governmental funders for the benefit of astronomy generally. Many of the partners in all of these telescopes have their own policies, practices, and legal requirements that must be taken into account.[13]

---

[12] The exceptions here are notable, and we can acknowledge the worry that non-collaborative states are more likely than others to act from realpolitik motivations.

This model of operation is strikingly different from, e.g., nuclear technology research, where government and corporate activities have been deeply intermeshed for decades, and where security has been a top priority since those fields' inception. Attempting to impose stringent security measures on not just the SETI community but the entire world's radio astronomy community now—after decades of flexible, multi-stakeholder collaboration—would require a sea change in how those activities operate, and it would run counter to the goals of many involved (that is, assuming the security requirements didn't have the effect of ending the projects entirely.)

Finally, it is important that implementing extensive security protections in the SETI and METI fields could itself cause the very problems W&T warn about. The existence of hardened facilities and locked-down information flows could itself be interpreted by outsiders as evidence that some world-altering activity was occurring within that community or facility, thus leading to exactly the kind of espionage and conflict that W&T are trying to avoid in the first place, even if nothing had actually been discovered.

Again, W&T's legitimate worry is that the mere perception of an information monopoly could be enough to generate dangerous conflict. So insofar as this worry takes precedence, we ought to avoid security measures that reinforce that perception. These recommendations are thus to some degree self-defeating.

However, even if we have good reason to avoid extensive security protections of facilities per se, there remain other reasons to enact security measures meant to protect the SETI practitioners themselves, especially in the event of detection. The scientists who conduct this work could easily become the targets of online or real-world harassment and attacks, to which we know many scientists have unfortunately already been subject.[14] And even though the equipment likely needed to communicate with ETI is common and the scientific information likely to be made widely available through collaboration, expertise will still be needed to conduct and interpret the work involved, which lends support to the claim that there ought to be protections in place for the scientists themselves who possess the relevant expertise.

---

[13] As a specific example of the sort of multi-stakeholder situation that describes much of astronomy and SETI: the Breakthrough Listen SETI Initiative is funded primarily by the Breakthrough Prize Foundation, founded by billionaire philanthropist Yuri Milner, who was born in Russia, is an Israeli citizen, and lives in the US. Its scientists are largely employees of the University of California, Berkeley and other astronomy-related institutions around the world, and its expenses largely consist of deploying specialized hardware at existing large national and private observatories, from which it purchases telescope observing time. The data it collects is stored on servers at various facilities including observatories, universities, and in the commercial cloud.

[14] E.g., https://alumni.berkeley.edu/california-magazine/summer-2020/michael-mann-on-climate-denial-and-doom , www.nature.com/articles/d41586-021-02741-x

# 7. Alternative policy options

The geopolitical dangers presented by W&T are not inevitable. It is true that W&T argue that even if their scenario is low probability, it is high impact and so must be prepared for, but as we have shown, there are other, competing contact and geopolitical scenarios that lead to different recommendations, and W&T do not give good reason to prioritize realpolitik risks above all others

The W&T realpolitik scenario is also less likely than W&T imply. While the potential misperception of an information monopoly is a real cause for concern, and W&T are wise to highlight it, the presumption of an actual information monopoly rests on a series of flawed assumptions. Regardless, policy tools are available that could be adopted by the SETI/METI community and policymakers more generally to mitigate the danger posed by the misperception of an information monopoly and any lingering realpolitik concerns.

Instead of attempting to control information relating to successful SETI/METI efforts and starting an arms race, the international community could adopt precisely the opposite approach: full, open, and transparent information sharing. Building on existing post-detection protocols, states could commit themselves to norms or even legally-binding rules that mandate open access to the results of SETI and METI activities. If followed, such an approach would eliminate the dangers described by W&T, because no monopoly on detection or communication would be possible.

Notably, a transparent approach will also mitigate any dangers posed by an erroneous perception on the part of states that someone else has achieved SETI or METI success. This could include the dissemination of information that will help to dispel the myth that, in the event of detection, an information power grab is a viable option, thereby providing motivation for collaboration that more competitive states might otherwise have been lacking. The solution, in this case, is not to recommend hardening radio telescope security. This security and secrecy could provoke misperceptions and promote panic rather than protect against those very dangers. The solution, instead, is to ensure SETI practitioners transparently share their data and educate government officials about their methods and findings, or lack thereof (but see [35] for a counterpoint).

There is precedent for such an approach. The 1967 Outer Space Treaty requires states to "inform the Secretary-General of the United Nations as well as the public and the international scientific community, to the greatest extent feasible and practicable, of the nature, conduct, locations and results of [activities in outer space]." (Article XI) More specifically, the Treaty makes stations, installations, equipment, and vehicles on the Moon and other celestial bodies open to inspections and visits by other parties to the Treaty, on a basis of reciprocity. (Article XII) Likewise, Article VII of the Antarctic Treaty allows Parties to that agreement to designate observers, who can at any time inspect "all areas of Antarctica, including all stations, installations, and equipment…" Although far from perfect, these examples demonstrate that the international community can and has used information sharing and transparency measures to prevent conflicts that might arise from the results of scientific activities.

Such an approach need not depend on the force of international law to be effective. Practical measures could be put in place to reinforce this system of data-sharing and transparency. For example, many hundreds of agreements between countries already exist that provide mechanisms for the exchange of scientific personnel and international collaboration on research projects. Using these mechanisms, states could ensure that trusted researchers are involved in SETI/METI activities undertaken by foreign counterparts. In addition, SETI/METI practitioners could adopt practices similar to those used by NASA, the European Space Agency, and others, whereby the raw data they collect is made publicly available as soon as technically possible, without any political intermediary, as is largely already being done by, for instance, the Breakthrough Listen Initiative [53]. Any deviation from this practice could then be detected by other practitioners, alerting the community that something might be being suppressed. A combination of these practical steps, backed by internationally-agreed norms and rules, would undermine the sort of race for control described by W&T and render the tasks listed in their recommendations unnecessary, while having the added benefit of enhancing SETI and METI efforts by easing international collaboration.

The education of government officials who might trigger W&T's concerns is also an essential antidote here. The W&T realpolitik scenario is a concern because it relies only on the perception that their contact scenario has occurred. The primary response should therefore be not to treat this perception as a foregone conclusion, but to dispel it. This would ideally occur as part of the work on post-detection protocols and information sharing described above, but would certainly also follow any actual contact scenario, potentially defusing the W&T realpolitik scenario shortly after it begins. This education would flow both ways, helping SETI practitioners understand the attitudes of those who might react badly in the face of a detection, and encourage them to take appropriate measures.

Finally, we note the limitations of any preparations for contact. While such preparations are prudent, the actual best way forward immediately after contact is established will strongly depend on the nature of the contact and the content of any signals detected. As explained in the Billingham report, quoting Kopal (1990) [54]:

> It would not be wise to request an immediate consideration of a special regulation governing the SETI/CETI activities…The consideration of a legislative process can only be started when the boundary between possibilities and well-established realities has been crossed.

and by Denning (2013) [55]:

> "any matter of policy can only be sensibly based on an acknowledgement that we do not know the risks of contact, rather than efforts to calculate the unimaginable…[and] that in the majority of imaginable detection or contact scenarios, people will not be dealing with the facts or other life forms, but rather, with our distorted and refracted and fractured cultural representations of those facts and entities…[and] that contact has now been rehearsed so many times in popular culture that these representations and their dissemination in new media will be influential beyond almost any other factor".

# 8. Conclusions

We agree that a realpolitik response to a contact scenario is worth considering, but we maintain that it is just one of the various candidate post-contact responses that merit consideration. We find reason to worry that W&T's presentation of realpolitik motivations is insufficient and overly simplistic, and we at the same time motivate the inclusion of other candidate responses to contact in our geopolitical considerations, specifically those that might generate cohesion or greater collaboration at the level of international relations. We motivate consideration of the possibility of a global collective that transcends national collectives. We consider how the collective interests of humanity might diverge from or conflict with W&T's presentation of the interests of national or governmental collectives, and we caution against the potential for the misrepresentation of purportedly collective interests in geopolitical relations. We also challenge W&T's claim that SETI is as risky as METI, highlighting fundamental features of METI that could increase the risk of a realpolitik response and the concomitant potential for conflict or catastrophe on which W&T focus.

We find that the justifications for policy recommendations in light of the W&T realpolitik scenario rest crucially on the possibility that politicians or other leaders will *misperceive* the potential for an information monopoly. We agree with W&T that this real potential for misperception merits certain policy recommendations, but we maintain that the appropriate recommendations differ significantly from those suggested by W&T. Furthermore, the likelihood of an *actual* information monopoly is low, and the justifications provided by W&T for the potential of a real information monopoly are based on W&T's contact scenario—an overly narrow and implausible scenario that is inconsistent with what our current science indicates and what some SETI practitioners expect to find. Ultimately, we find that the majority of W&T's recommendations are unlikely to prevent the ills they foresee, and that they may even precipitate them by fostering the aforementioned misperception that a government has already achieved a monopoly on strategically useful information.

We agree, at least, that there is good reason to apply post-contact security measures for the SETI practitioners themselves whose expertise might place them at greater risk in any contact scenario. As an alternative to W&T's other recommendations (e.g. that sites of SETI research, such as radio telescopes, have their security hardened), we instead recommend transparency, data sharing, and education of policymakers. Specifically, we recommend that governments, learned societies, and international organizations such as the International Academy of

Astronautics continue the unfinished work of the Billingham workshops to establish norms and standards for educating policymakers and for post-detection protocols. Some of this work already occurs in, for example, briefings to Congress, frequent articles in popular science presses, and recurring AAAS discussions, in addition to the academic discourse we discuss in Section 1.4. The report itself recommends that:

> Government decision-makers in general should be informed about SETI research and its implications so that they are better prepared to anticipate and deal with subsequent events if there is a signal detection. Means of doing this would include briefings for selected officials and legislators and the broad distribution of publications on SETI and its implications.

and

> At the international level…agencies involved in SETI research should inform officials of appropriate international organizations, and national representatives to those organizations, about SETI and its possible outcomes.

# Acknowledgements

The Penn State Extraterrestrial Intelligence Center and the Center for Exoplanets and Habitable Worlds are supported by the Eberly College of Science and the Pennsylvania State University. The views expressed in this article are those of the authors, and do not necessarily represent those of their employers. We are grateful to Kathryn Denning and Thomas Metcalf for their detailed comments on earlier drafts. We are also grateful to our referees for their valuable feedback.

# References

[1]     G. Cocconi and P. Morrison. Searching for Interstellar Communications. Nature, 184:844–846, September 1959. doi: 10.1038/184844a0.
[2]     F. D. Drake. Project OZMA. Physics Today, 14:40–46, April 1961. doi: 10.1063/1.3128863.
[3]     J. C. Tarter. The evolution of life in the Universe: are we alone? Highlights of Astronomy, 14:14–29, August 2007. doi: 10.1017/S1743921307009829.
[4]     Jason T. Wright, Sofia Sheikh, Iván Almár, Kathryn Denning, Steven Dick, and Jill Tarter. Recommendations from the Ad Hoc Committee on SETI Nomenclature. arXiv e-prints, art. arXiv:1809.06857, Sep 2018.

[5]     Staff at the National Astronomy and Ionosphere Center. The Arecibo message of November, 1974. Icarus, 26(4):462–466, December 1975. doi: 10.1016/0019-1035(75)90116-5.

[6]     Alexander Zaitsev. Messaging to Extra-Terrestrial Intelligence. arXiv e-prints, art. physics/0610031, October 2006.

[7]     William Macauley. Science From Beyond: NASA's Pioneer Plaque and the History of Interstellar Communication, 1957–1972. In Workshop on Extraterrestrial life—Beyond our expectations?, page 26, May 2012.

[8]     J. M. Pasachoff. History of Astronomy: The Voyager Golden Record in Intellectual History. In AAS/Division for Planetary Sciences Meeting Abstracts, volume 52 of AAS/Division for Planetary Sciences Meeting Abstracts, page 407.01, October 2020.

[9]     Douglas A. Vakoch. Asymmetry in Active SETI: A case for transmissions from Earth. Acta Astronautica, 68(3):476–488, February 2011. doi: 10.1016/j.actaastro.2010.03.008.

[10]    USSR Academy of Sciences. The Soviet CETI Program. Icarus, 26(3): 377–385, November 1975. doi: 10.1016/0019-1035(75)90182-7.

[11]    Alexander Zaitsev. The SETI Paradox. arXiv:physics/0611283, art. physics/0611283, November 2006.

[12]    T. Cortellesi. Reworking the SETI Paradox: METI's Place on the Continuum of Astrobiological Signaling. arXiv:2006.01167, art. arXiv:2006.01167, June 2020.

[13]    Sofia Z. Sheikh, Andrew Siemion, J. Emilio Enriquez, Danny C. Price, Howard Isaacson, Matt Lebofsky, Vishal Gajjar, and Paul Kalas. The Breakthrough Listen Search for Intelligent Life: A 3.95-8.00 GHz Search for Radio Technosignatures in the Restricted Earth Transit Zone. AJ, 160 (1):29, July 2020. doi: 10.3847/1538-3881/ab9361.

[14]    J. Billingham. Social Implications of the Detection of an Extraterrestrial Civilization: A Report of the Workshops on the Cultural Aspects of SETI Held in Three Sessions, October 1991, May 1992, and September 1992 at Santa Cruz, California. SETI Press, 1999. ISBN 9780966633504. URL https://books.google.com/books?id=iLUrAAAAYAAJ.

[15]    Jason T. Wright and Michael P. Oman-Reagan. Visions of human futures in space and SETI. International Journal of Astrobiology, 17(2):177–188, April 2018. doi: 10.1017/S1473550417000222.

[16]    Jill Tarter. Contact who will speak for earth and should they? In C. Impey, A.H. Spitz, and W. Stoeger, editors, Encountering Life in the Universe: Ethical Foundations and Social Implications of Astrobiology. University of Arizona Press, 2013. ISBN 9780816528707. URL https://books.google.com/books?id=LUt5AAAAQBAJ.

[17]    Chelsea Haramia and Julia DeMarines. The imperative to develop an ethically-informed meti analysis. Theology and Science, 17:38–48, 01 2019. doi: 10.1080/14746700.2019.1557800.

[18]    John W. Traphagan. Should We Lie to Extraterrestrials? A Critique of the Voyager Golden Records. Space Policy, 57:101440, August 2021. doi: 10.1016/j.spacepol.2021.101440.

[19]    S. Shostak. Are transmissions to space dangerous? International Journal of Astrobiology, 12:17–20, January 2013. doi: 10.1017/S1473550412000274.

[20]    Jacob Haqq-Misra, Michael W. Busch, Sanjoy M. Som, and Seth D. Baum. The benefits and harm of transmitting into space. Space Policy, 29(1): 40–48, February 2013. doi: 10.1016/j.spacepol.2012.11.006.

[21]     Alan Penny.   Transmitting (and listening) may be good  (or bad). Acta Astronautica, 78:69–71, September 2012. doi: 10.1016/j.actaastro.2011.09.013.

[22]     S. D. Baum, J. D. Haqq-Misra, and S. D. Domagal-Goldman. Would contact with extraterrestrials benefit or harm humanity?  A scenario analysis. Acta Astronautica, 68:2114–2129, June 2011. doi: 10.1016/j.actaastro.2010.10.012.

[23]     Jerome H Barkow. Eliciting altruism while avoiding xenophobia: A thought experiment. In Extraterrestrial Altruism, pages 37–48. Springer, 2014.

[24]     Carlos Santana. We Come in Peace? A Rational Approach to METI. Space Policy, 57:101430, August 2021. doi: 10.1016/j.spacepol.2021.101430.

[25]     J. Billingham and James Benford. Costs and Difficulties of Interstellar 'Messaging' and the Need for International Debate on Potential Risks. Journal of the British Interplanetary Society, 67:17–23, January 2014.

[26]     J. Gertz. Reviewing METI: A Critical Analysis of the Arguments. ArXiv e-prints, May 2016.

[27]     Iván Almár and Jill Tarter.  The discovery of ETI as a high-consequence, low-probability event. Acta Astronautica, 68(3):358–361, February 2011. doi: 10.1016/j.actaastro.2009.07.007.

[28]     Kathryn Denning and Steven J. Dick. Preparing for the Discovery of Life Beyond Earth. In Bulletin of the American Astronomical Society, volume 51, page 183, September 2019.

[29]     Martin Dominik and John C. Zarnecki. The detection of extra-terrestrial life and the consequences for science and society. Philosophical Transactions of the Royal Society. A, Mathematical, Physical and Engineering Sciences, 369(1936):499–507, February 2011. ISSN 1364-503X. doi: 10.1098/rsta.2010.0236.

[30]     Steven J. Dick. Astrobiology and Society: An Overview. In K. C. Smith and
C. Mariscal, editors, Social and Conceptual Issues in Astrobiology, page 9. 2020.

[31]     Kenneth W. Wisian and John W. Traphagan. The Search for Extraterrestrial Intelligence: A Realpolitik Consideration. Space Policy, 52:101377, May 2020. doi: 10.1016/j.spacepol.2020.101377.

[32]     John Bew. Realpolitik: a history. Oxford University Press, 2016.

[33]     J. E. Enriquez, A. Siemion, G. Foster, V. Gajjar, G. Hellbourg, J. Hickish, H. Isaacson, D. C. Price, S. Croft, D. DeBoer, M. Lebofsky, D. H. E. MacMahon, and D. Werthimer. The Breakthrough Listen Search for Intelligent Life: 1.1-1.9 GHz Observations of 692 Nearby Stars. ApJ, 849:104, November 2017. doi: 10.3847/1538-4357/aa8d1b.

[34]     R. S. Dixon. A Search Strategy for Finding Extraterrestrial Radio Beacons. Icarus, 20:187–199, October 1973. doi: 10.1016/0019-1035(73)90050-X.

[35]     J. Gertz.  Post-Detection SETI Protocols & METI: The Time Has Come To Regulate Them Both. ArXiv e-prints, January 2017.

[36]     Shelley A. Wright, Paul Horowitz, Jérôme Maire, Dan Werthimer, Franklin Antonio, Michael Aronson, Sam Chaim-Weismann, Maren Cosens, Frank D. Drake, Andrew W. Howard, Geoffrey W. Marcy, Rick Raffanti, Andrew P. V. Siemion, Remington P. S. Stone, Richard R. Treffers, and Avinash Uttamchandani. Panoramic optical and near-infrared SETI instrument: overall specifications and science program. In Ground-based and Airborne Instrumentation for Astronomy VII, volume 10702 of Society of Photo-Optical Instrumentation Engineers (SPIE) Conference Series, page 107025I, Jul 2018. doi: 10.1117/12.2314268.

[37] C. Sagan. Contact: a novel. Simon and Schuster, 1985. ISBN 9780671434007. URL https://books.google.com/books?id=ESzHK3zvwaMC.

[38] Steve Starkey and Robert Zemeckis. Contact. Warner Bros., USA, 1997.

[39] Ted Chiang. Story of your life. In Starflight 2. Tor Books, New York, New York, USA, 1998.

[40] Shawn Levy, Dan Levine, Aaron Ryder, and David Linde. Arrival. Paramount Pictures, USA, 2016.

[41] H.J. Morgenthau. Politics Among Nations: The Struggle for Power and Peace. Borzoi book. A. A. Knopf, 1948. URL https://books.google.com/books?id=M9IlAAAAMAAJ.

[42] K.N. Waltz. Theory of International Politics. Addison-Wesley series in political science. McGraw-Hill, 1979. ISBN 9780075548522. URL https://books.google.com/books?id=j6qOAAAAMAAJ.

[43] J.J. Mearsheimer, G. Alterman, and R.W.H.D.S.P.P.S.C.D.P.I.S.P.J.J. Mearsheimer. The Tragedy of Great Power Politics. Monologue audition series. Norton, 2001. ISBN 9780393020250. URL https://books.google.com/books?id=jOV9HuCppqwC.

[44] W.N.W. Cobb. Privatizing Peace: How Commerce Can Reduce Conflict in Space. Taylor & Francis, 2020. ISBN 9781000095425. URL https://books.google.com/books?id=YbvrDwAAQBAJ.

[45] Dimitrios Stroikos. Engineering world society? scientists, internationalism, and the advent of the space age. International Politics, 55(1):73–90, 2018.

[46] N. Bormann and M. Sheehan. Securing Outer Space: International Relations Theory and the Politics of Space. Routledge Critical Security Studies. Taylor & Francis, 2009. ISBN 9780203882023. URL https://books.google.com/books?id=XM85r5dmWlIC.

[47] Yuen Foong Khong. Power as prestige in world politics. International Affairs, 95(1):119–142, 01 2019. ISSN 0020-5850. doi: 10.1093/ia/iiy245. URL https://doi.org/10.1093/ia/iiy245.

[48] Terrance Allen. China's realist dilemma : the pursuit of international prestige. PhD thesis, Air University (US) School of Advanced Air And Space Studies, 2018.

[49] Yanzhong Huang. Vaccine diplomacy is paying off for china. 2021.

[50] Moritz Rudolf. China's health diplomacy during Covid-19: the Belt and Road Initiative (BRI) in action, volume 9/2021 of SWP Comment. Stiftung Wissenschaft und Politik -SWP- Deutsches Institut für Internationale Politik und Sicherheit, Berlin, 2021. doi: https://doi.org/10.18449/2021C09.

[51] Michael Huemer. The Problem of Political Authority, pages 3–19. Palgrave Macmillan UK, London, 2013. ISBN 978-1-137-28166-1. doi:10.1057/9781137281661 1. URL https://doi.org/10.1086/673423

[52] David Robert Grimes. On the viability of conspiratorial beliefs. PloS one, 11(1):e0147905, 2016.

[53] Matthew Lebofsky, Steve Croft, Andrew P. V. Siemion, Danny C. Price, J. Emilio Enriquez, Howard Isaacson, David H. E. MacMahon, David Anderson, Bryan Brzycki, Jeff Cobb, Daniel Czech, David DeBoer, Julia DeMarines, Jamie Drew, Griffin Foster, Vishal Gajjar, Nectaria Gizani, Greg Hellbourg, Eric J. Korpela, Brian Lacki, Sofia Sheikh, Dan Werthimer, Pete Worden, Alex Yu, and Yunfan Gerry Zhang. The Breakthrough Listen Search

for Intelligent Life: Public Data, Formats, Reduction, and Archiving. PASP, 131(1006):124505, December 2019. doi: 10.1088/1538-3873/ab3e82.

[54] Vladimir Kopal. International law implications of the detection of extraterrestrial intelligent signals. Acta Astronautica, 21(2):123–126, January 1990. doi: 10.1016/0094-5765(90)90138-B.

[55] Kathryn Denning. Impossible predictions of the unprecedented: Analogy, history, and the work of prognostication. In Douglas A. Vakoch, editor, Astrobiology, History, and Society, Advances in As- trobiology and Biogeophysics, pages 301–312. Springer Berlin Heidelberg, 2013. ISBN 978-3-642-35982-8 978-3-642-35983-5. URL https://link.springer.com/chapter/10.1007%2F978-3-642-35983-5_16#page-1